# Terahertz switching of antiferromagnetic order by Néel spin-orbit torques

Y. Behovits[1], A. L. Chekhov[1], A. Ruge[1], R. Rouzegar[1], B. Rosinus Serrano[1], A. P. Fellows[2], B. John[2], M. Thämer[2], S. Reimers[3], F. Renner[4], T. Dannegger[4], U. Nowak[4], T.S. Seifert[1], M. Kläui[3], M. Jourdan[3] and T. Kampfrath[1]

1. Department of Physics, Freie Universität Berlin, 14195 Berlin, Germany
2. Department of Physical Chemistry, Fritz-Haber Institute of the Max Planck Society, 14195 Berlin, Germany
3. Institute of Physics, Johannes Gutenberg-Universität Mainz, 55099 Mainz, Germany
4. Department of Physics and Center for Applied Photonics, University of Konstanz, 78457 Konstanz, Germany

**Summary**

Ultrafast electric manipulation of magnetic order in solids is critical for the development of future terahertz data processing [1, 2]. A fascinating concept for such high-speed operation is offered in metallic antiferromagnets by Néel spin-orbit torque [3, 4]. It should allow one to coherently rotate the ordered spins by simply applying an electric current of suitable amplitude and polarity [5]. However, such switching has been severely hampered by competing heat-induced effects [6-8], and it has not yet been achieved on the intrinsically ultrafast time scales of antiferromagnets. Here, we report robust, direction-controlled and non-thermal rotation of the Néel vector $L$ by $\pm 90°$ at room temperature in the antiferromagnet $Mn_2Au$ driven by phase-locked terahertz current pulses. All observed features are consistent with ultrafast Néel spin-orbit torque: First, nonlinear optical imaging reveals that the terahertz current direction sets the final orientation of $L$ in the absence of any bias field for at least two months. Second, transient optical birefringence shows that the switching proceeds ultrafast in less than 15 picoseconds. Finally, atomistic spin-dynamics simulations reproduce the observed dynamics and confirm the minor role of thermal effects. While the switching is already one order of magnitude faster than in ferromagnets at comparable dissipated energy [9], our simulations predict routes toward switching times and energies which are another order of magnitude lower. Our approach can be transferred to electric-field-driven switching in many more antiferromagnets [10-16], including magnetoelectric insulators [17-19]. The engineering of spin torques, resonance frequencies and read-out mechanisms provides an exciting pathway toward on-chip applications of terahertz antiferromagnetic spin-orbitronics.

**Figures**

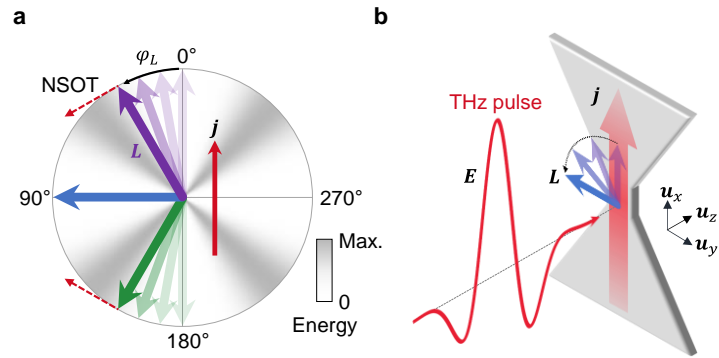

**Fig. 1 | Concept of ultrafast Néel spin-orbit torque. a**, Schematic top view on the plane of a thin film of the metallic antiferromagnet Mn₂Au. The Néel vector $L$ has four stable orientations with angle $\varphi_L = 0°, 90°, 180°, 270°$ in the anisotropy energy landscape, as indicated by the white-gray color map. A charge current with density $j$ (red arrow) induces field-like Néel spin-orbit torque (NSOT) that accelerates the Néel vector by $\partial_t^2 \varphi_L \propto L \cdot j$. Consequently, $j$ drives domains with an initial Néel vector of $+L_0$ ($\varphi_{L_0} = 0°$, purple) and $-L_0$ ($\varphi_{L_0} = 180°$, green) toward the same minimum at $\varphi_L = 90°$ (blue). **b**, Approach to ultrafast NSOT. An incident terahertz electric-field pulse $E$ (red) drives a charge current $j$ (red) inside a Mn₂Au thin film and, thus, exerts NSOT on the Néel vector $L$ (blue). To enhance the charge current locally, the film is microstructured in the shape of a bowtie antenna.

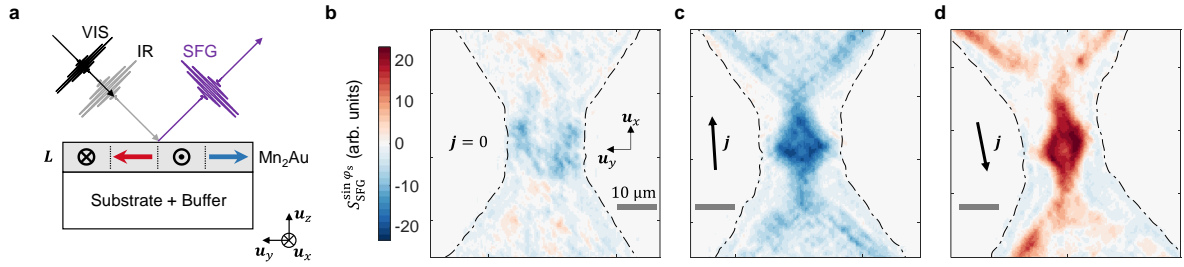

**Fig. 2 | Micrographs of the Néel vector. a**, Schematic of sum-frequency-generation (SFG) probing and sample structure. Incident infrared (IR) and visible (VIS) pulses (p-polarized) interact in Mn$_2$Au. They generate a IR+VIS sum-frequency response that reports on the local net Néel vector $\langle L \rangle$ with a spatial resolution of ~ 1 μm. The four equilibrium $L$ orientations are indicated by arrows. **b**, Micrograph of the SFG signal $S_{\mathrm{SFG}}^{\sin\varphi_{\mathrm{S}}}$ from an unexposed reference antenna, **c**, an antenna exposed to $10^4$ terahertz charge-current pulses $j$ with positive polarity and **d**, an antenna exposed to $10^4$ pulses with negative polarity. The signal $S_{\mathrm{SFG}}^{\sin\varphi_{\mathrm{S}}}$ primarily scales with the net Néel-vector component $\langle L_y \rangle$ [Eq. (2)], is normalized by a reference signal (see Methods) and color-coded according to the color bar. The complementary signal reporting on $\langle L_x \rangle$ is shown in Supplementary Fig. 4.

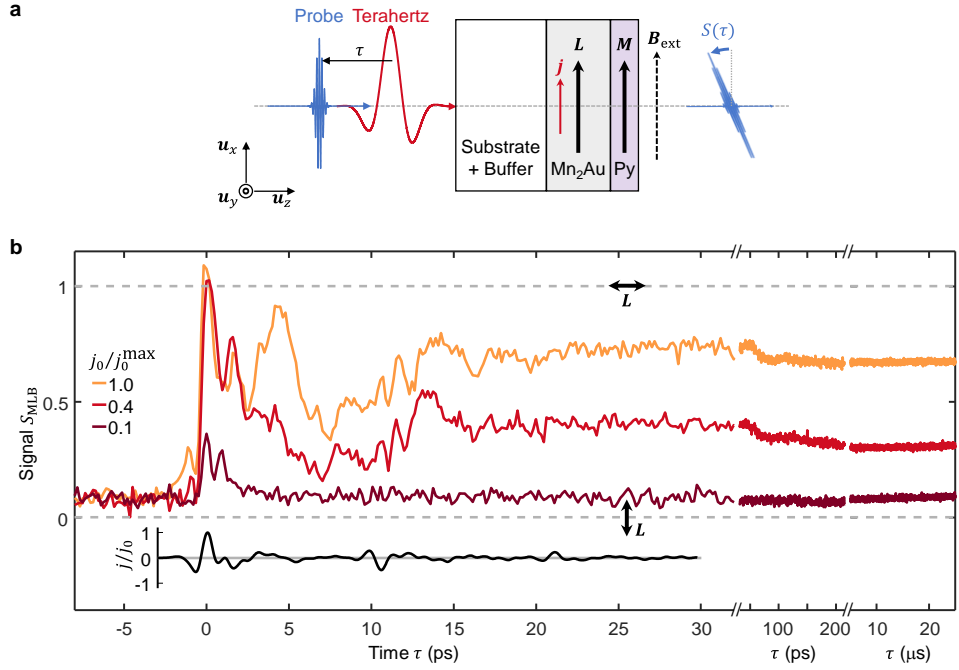

**Fig. 3 | Terahertz Néel-vector dynamics. a**, Schematic of the terahertz-pump magneto-optic-probe experiment with terahertz current $\boldsymbol{j} = j\boldsymbol{u}_x$. Prior to the terahertz pump pulse, an external magnetic field $\boldsymbol{B}_{\text{ext}} = B_{\text{ext}}\boldsymbol{u}_x$ sets the magnetization $\boldsymbol{M}$ of the ferromagnetic Ni$_{80}$Fe$_{20}$-layer (Py) adjacent to the antiferromagnetic Mn$_2$Au and, thus, initializes its Néel vector $\boldsymbol{L}$. At a delay $\tau$ after the pump, the optical probe pulse interrogates $\boldsymbol{L}$ through magnetic linear birefringence (MLB). **b**, MLB signals $S_{\text{MLB}}(\tau)$ vs $\tau$ [Eq. (3)] for a peak current density $j_0 = \max|j(t)|$ of $j_0/j_0^{\max} = 0.1$ (yellow), 0.4 (orange) and 1.0 (red), where $j_0^{\max} \approx 3 \cdot 10^{13}$ A/m$^2$. The black curve displays the reconstructed terahertz current density $j(\tau)/j_0$ inside the Mn$_2$Au film. As the signals predominantly scale with $\langle \sin^2 \varphi_L \rangle$ [Eq. (3)], signal values of 0 and 1 (gray-dashed lines) correspond to fully horizontally and vertically aligned Néel vectors within the probing volume (see black double-arrows).

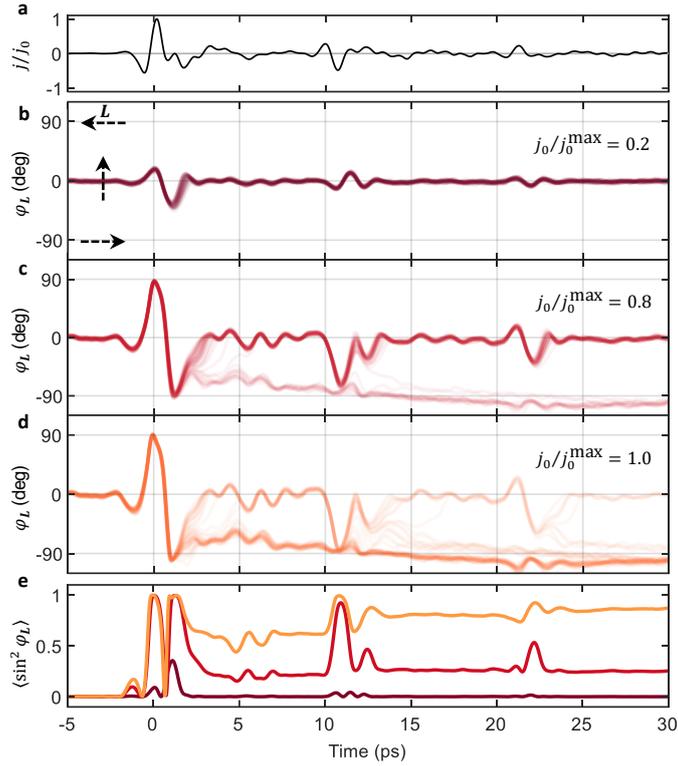

**Fig. 4 | Atomistic spin-dynamics simulations. a**, Waveform of the terahertz current density $j(\tau)/j_0$. **b-d**, Simulated Néel-vector deflection $\varphi_L(\tau)$ for increasing terahertz peak current density $j_0/j_0^{\max}$ at a thermostat temperature of 300 K. The $\varphi_L(\tau)$ is averaged over the 30 nm thickness of the Mn$_2$Au thin film. The impact of thermal, stochastic fields is assessed by repeating each simulation 50 times. The individual results are shown by semi-transparent lines. Black arrows in panel b show the corresponding direction of the Néel vector $L$. **e**, Calculated MLB signals $\langle \sin^2 \varphi_L \rangle$, obtained by averaging over the 50 simulation runs shown in each panel b-d.

**Main text**

**Introduction**

Antiferromagnets have powerful benefits over ferromagnets with regard to future applications in ultrafast data processing [1, 2]. Examples include smaller volume for storing one bit and robustness against perturbing magnetic fields up to $10\,\mathrm{T}$. Importantly, long-wavelength magnons at extremely high frequencies of $\sim 1\,\mathrm{THz}$ permit ultrafast uniform coherent spin dynamics and even magnetic switching [20-22]. Further, antiferromagnets are abundant and come with various spatial symmetries that enable novel functionalities [23, 24].

A fascinating example is the field-like Néel spin-orbit torque (NSOT) that emerges when collinear antiferromagnetic order breaks the inversion symmetry of the crystal lattice [3]. In the metallic antiferromagnets CuMnAs and Mn$_2$Au, this NSOT can simply be induced by electric currents (Fig. 1a). Ultimately, currents of suitable amplitude and polarity should allow one to rotate the Néel vector $\boldsymbol{L}$ by $\pm 90°$ between different equilibrium configurations [4, 25-28]. Owing to exchange enhancement, such switching is predicted to be efficient and ultrafast down to the picosecond scale, which is the natural time scale of antiferromagnetic spin dynamics [5, 29, 30].

However, ultrafast implementations of genuine NSOT switching are missing because slow, quasi-DC switching is extremely challenging already: (1) The required current density of $\gtrsim 10^{11}\,\mathrm{A/m^2}$ induces competing processes, such as heating, strain, domain fragmentation and structural changes, which dominate for current pulses longer than $\sim 10\,\mathrm{ns}$ [6-8, 31, 32]. (2) Exclusive probing of antiferromagnetic order is nontrivial, in particular the differentiation of antiparallel states $\pm \boldsymbol{L}$ [33].

To address challenge (1), phase-locked sub-picosecond current transients are ideally suited (Fig. 1b): Due to their terahertz center frequency, they can resonantly and, thus, efficiently excite coherent antiferromagnetic magnons [34-38]. The sub-picosecond duration is mandatory for ultrafast switching and reduces the dissipated heat relative to longer pulses [8, 39, 40]. To tackle challenge (2), optical probes are a highly versatile tool: Magnetic linear birefringence (MLB) and dichroism can differentiate 0°- and 90°-domains [33], and second-order nonlinear optical effects, such as optical sum-frequency generation (SFG), offer complementary information on $\pm \boldsymbol{L}$ states [41-43].

Here, we report the long-sought ultrafast switching of the Néel vector $\boldsymbol{L}$ of the metallic collinear antiferromagnet Mn$_2$Au at room temperature by terahertz NSOT (Fig. 1b). First, in Mn$_2$Au thin films, SFG microscopy reveals signatures of direction-controlled $\pm 90°$-switching of $\boldsymbol{L}$ without bias field, which persist over at least two months [44, 45]. Second, in Mn$_2$Au|Py stacks with the ferromagnet Py, where $\boldsymbol{L}$ can easily be initialized, we time-resolve the switching process by MLB [46] over a time window from $-10\,\mathrm{ps}$ to as high as $+1\,\mathrm{ms}$ with femtosecond resolution. We find that ultrafast rotation of $\boldsymbol{L}$ by $90°$ happens in less than $15\,\mathrm{ps}$ and remains stable until the sample is externally reinitialized after $\sim 0.1\,\mathrm{ms}$. Our approach takes full advantage of the unique features of NSOT and antiferromagnets: direction-controlled, non-thermal, robust, ultrafast and energy-efficient. We expect that it can be transferred to many more antiferromagnets and, thus, be considered a major step toward terahertz antiferromagnetic spin-orbitronics.

**Principle of NSOT-driven 90°-switching**

Fig. 1a schematically shows the plane of a Mn$_2$Au thin film. The Néel vector $\boldsymbol{L}$ is given by the difference of the magnetization of the two spin sublattices, which are antiparallel in equilibrium. In a given domain, $\boldsymbol{L}$ can take one out of four $90°$-rotated orientations in the magnetic potential energy landscape that arises from various anisotropy interactions. Remarkably, by applying an electric current with density $\boldsymbol{j}$ in the film plane (Fig. 1a), NSOT is induced and initiates rotation of the Néel vector according to [46, 47]

$$\partial_t^2 \varphi_L |_{\mathrm{NSOT}} \propto \boldsymbol{L} \cdot \boldsymbol{j} = |\boldsymbol{L}||\boldsymbol{j}|\cos(\varphi_L - \varphi_j). \qquad (1)$$

If the NSOT is sufficiently strong, it pushes the Néel vector over the energy barrier (gray-shaded areas), thereby achieving $90°$-switching from the initial $\boldsymbol{L}_0$ at $\varphi_{L_0} = 0°$ to the new equilibrium at $\varphi_L = 90°$ (Fig. 1a).

The NSOT of Eq. (1) has a unique feature: It reverses when either $\boldsymbol{L}$ or $\boldsymbol{j}$ is reversed. Therefore, if we have two domains with opposite initial Néel vectors of $\pm \boldsymbol{L}_0$ ($\varphi_{L_0} = 0°$ and $180°$) and apply $\boldsymbol{j}$ parallel to $L_0$, both domains can be switched toward the same final direction $\varphi_{L_0} = 90°$ (Fig. 1a). Consequently, NSOT permits macroscopic alignment of the Néel vector in samples that are initially multi-domain, thereby turning a sample with vanishing volume-averaged Néel vector $\langle L \rangle$ into a sample with $\langle L \rangle \neq 0$, without any external bias. Likewise, opposite final $\pm \langle L \rangle$ are obtained for opposite current polarities $\pm j$ and, thus, opposite driving electric fields $\pm E$ [Eq. (1)].

In contrast, in centrosymmetric antiferromagnets, NSOT and the more general magneto-electric effect are symmetry-forbidden [18, 19, 48]. Consequently, the reorientation of $L$ is exclusively even in $E$ [20, 21] and, for certain spin modes, also even in the driving magnetic field [22]. It is, thus, independent of the polarity of the applied electromagnetic field. However, even if NSOT is symmetry-allowed, it is easily overwhelmed by polarity-independent Joule heating ($\propto E^2$) [6-8].

**Experiments**

To realize ultrafast NSOT switching, phase-locked terahertz pulses with peak electric fields up to $1.1\,\mathrm{MV/cm}$ (Supplementary Fig. 1 and Methods) are applied to an antenna-shaped Mn$_2$Au sample to drive a charge current $j$ and, thus, NSOT (Fig. 1b). Our samples consist of Mn$_2$Au thin films (thickness $30$-$40\,\mathrm{nm}$) grown on MgO with metallic buffer and cap layers. Each metal stack is laterally microstructured into a bowtie antenna (length $160\,\mathrm{\mu m}$, waist $20\,\mathrm{\mu m}$) to enhance the local current density by a factor of 5-15 (Supplementary Fig. 2, Supplementary Fig. 3). The antenna resonance frequency is chosen to overlap with the broadband response of the in-plane magnon mode of Mn$_2$Au at $0.6\,\mathrm{THz}$ [46].

The pump's impact on the antiferromagnetic order is probed by two magneto-optic approaches. (1) Long after excitation, we use phase-resolved optical SFG microscopy to acquire images of the $L$ landscape of the Mn$_2$Au thin film (Fig. 2a). (2) To time-resolve the spin dynamics in the antenna center, we use a time-delayed optical probe pulse and MLB (Fig. 3a).

**Long-term switching of Mn$_2$Au**

**Setup.** In the first experiment, we expose Mn$_2$Au antennas to $10^4$ consecutive terahertz pump pulses at a repetition rate of $1\,\mathrm{kHz}$. The terahertz current $j$ is parallel to the $x$ axis (Fig. 1b) with tunable polarity ($\pm j$).

Two months after exposure, the $L$ distribution of the sample is interrogated by wide-field SFG microscopy [49, 50]. As schematically shown in Fig. 2a, infrared (IR) and visible (VIS) laser pulses are collinearly incident onto the sample surface and generate an additional reflected wave at the IR+VIS sum frequency. The SFG field relies on the broken inversion symmetry of the sample, which, in bulk Mn$_2$Au, solely arises from the antiferromagnetic order. By taking SFG micrographs as a function of the rotation angle $\varphi_s$ of the sample about its normal, we extract two SFG signals that scale with $\cos\varphi_s$ and $\sin\varphi_s$ (Methods). They can be shown to fulfill

$$S_{\mathrm{SFG}}^{\cos\varphi_s} \propto -\langle L_x \rangle, \qquad S_{\mathrm{SFG}}^{\sin\varphi_s} \propto \langle L_y \rangle \qquad (2)$$

and should, thus, directly report on the in-plane Néel vector $\langle L \rangle$. Here, the brackets $\langle \cdot \rangle$ indicate averaging over the spatial resolution of $\sim 1\,\mathrm{\mu m}$. Importantly, $S_{\mathrm{SFG}}^{\cos\varphi_s}$ and $S_{\mathrm{SFG}}^{\sin\varphi_s}$ distinguish antiparallel states $\pm L$.

**Results.** Fig. 2b-d shows micrographs of $S_{\mathrm{SFG}}^{\sin\varphi_s}$ for three distinct antennas: one reference antenna without terahertz-pulse exposure (Fig. 2b), as well as antennas exposed to $10^4$ pulses with positive (Fig. 2c) or negative maximum amplitude (Fig. 2d). For the unexposed antenna, a random signal distribution with relatively small amplitudes is found (Fig. 2b). Toward the edges of the narrow antenna center, slightly increased SFG signals are found, which may indicate a modified magnetic anisotropy in these regions [51].

Remarkably, for the antenna exposed to terahertz pulses, a strong SFG signal in the antenna center emerges (Fig. 2c). Even more strikingly, for the sample exposed to terahertz pulses of opposite polarity, we again observe the emergence of a strong signal, but this time with opposite sign (Fig. 2d). The highest signal amplitudes form a characteristic shape, with a central diamond-shaped region and straight lines roughly following the edges of the widening antenna. These regions correspond to the sections with nearly identical current density, as the simulations in Supplementary Fig. 2 show. In contrast, we find little differences in the signal $S_{\mathrm{SFG}}^{\cos\varphi_s}$ between the three antennas (Supplementary Fig. 4).

Following Eq. (2), we interpret the SFG signals of Fig. 2b-d as images of the spatially averaged Néel vector $\langle L \rangle$, resulting in four important implications: (1) The terahertz pulse induces a nonvanishing macroscopic Néel vector $\langle L \rangle = \langle L_y \rangle u_y$ from an initial state with $\langle L_0 \rangle \approx 0$. (2) The induced $\langle L \rangle$ is perpendicular to the terahertz current density $j$. (3) $\langle L \rangle$ reverses when the polarity of the terahertz current $j$ is reversed. (4) The terahertz-induced $\langle L \rangle$ is stable for at least two months.

Results (1)-(3) are fully consistent with non-thermal switching through NSOT [Fig. 1b and Eq. (1)]. Thermal switching by Joule heating would be insensitive to the sign of the driving current, in contrast to Fig. 2c,d. Result (4) demonstrates the robustness of the switched antiferromagnetic textures.

**Terahertz switching dynamics in Mn$_2$Au|Py**

**Setup.** We now turn to the question whether the switching of the Néel vector occurs ultrafast and address it by a terahertz-pump optical-probe experiment (Fig. 3a). Because such a stroboscopic scheme requires many repetitions of the switching event, the sample needs to be initialized periodically. For this purpose, we use a Mn$_2$Au thin film (thickness 30 nm) that is exchange-coupled to an adjacent ferromagnetic-metal layer of Ni$_{80}$Fe$_{20}$ (Permalloy Py, 6 nm). It allows us to simultaneously control the uniform magnetization $M$ of the ferromagnet and the Néel vector $L \parallel M$ of the antiferromagnet by a moderate external magnetic field $B_{\text{ext}} = B_{\text{ext}} u_x$ [52, 53].

In each cycle of the experiment, we first apply $B_{\text{ext}} = +220$ mT to set the uniform Py magnetization and, thus, the Mn$_2$Au Néel vector $L$. By reducing $B_{\text{ext}}$ to zero, we obtain an approximately homogeneous distribution $L(x, y) = +L_0$. In this initialized remnant state, the incident terahertz pump pulse launches a current $j = ju_x$ and, thus, NSOT (Fig. 3a). After a delay $\tau$, the resulting $L$ dynamics is interrogated by a time-delayed optical probe pulse (duration 20 fs, wavelength 800 nm). The resulting signal $S(\tau, +L_0)$ is proportional to the probe's polarization rotation. Subsequently, we initialize $L(x, y) = -L_0$ and obtain the signal $S(\tau, -L_0)$. This procedure extends over 2 ms and is repeated periodically to sample $S(\tau, \pm L_0)$ at all relevant delays $\tau$ from $-10$ ps to as high as $+1$ ms.

The black curve in Fig. 3b shows the waveform of the current density $j(\tau)$ in the center of the antenna. The peak value $j_0 = \max|j(\tau)|$ can be set up to $j_0^{\max} \sim 3 \cdot 10^{13}$ A/m$^2$. The main pulse at $\tau \approx 0$ ps is followed by echoes at about 10 ps and 20 ps, which result from reflections at the substrate interfaces.

We now focus on the MLB signal component $S_{\text{MLB}}(\tau)$ that is even under reversal of $L_0$ (Methods). Calibration by the quasi-static sample response allows us to relate $S_{\text{MLB}}$ to the Néel-vector angle $\varphi_L$ by

$$S_{\text{MLB}}(\tau) = \frac{S(\tau, +L_0) + S(\tau, -L_0)}{2} \approx \langle \sin^2 \varphi_L(\tau) \rangle. \qquad (3)$$

Here, the brackets $\langle \cdot \rangle$ denote averaging over the probed Mn$_2$Au|Py area of $\approx 20 \times 30$ μm$^2$. According to Eq. (3), $S_{\text{MLB}}(\tau)$ distinguishes 90°-orientations of $L$, but is not sensitive to $\pm L$.

**Results.** Fig. 3b shows the signal $S_{\text{MLB}}(\tau)$ for increasing current amplitude inside the antenna. For the lowest peak amplitude $j_0 = 0.1 j_0^{\max}$, the waveform resembles the signals in Ref. [53]. In this approximately linear excitation regime, Eq. (3) implies $S_{\text{MLB}}(\tau) \propto \langle \varphi_L^2(\tau) \rangle$, which is consistent with the dynamics of a complimentary signal $\propto \langle \varphi_L(\tau) \rangle$ (Supplementary Fig. 5). Importantly, $S_{\text{MLB}}(\tau)$ recovers to its initial value within 5 ps and remains there until at least $\tau = 20$ μs (Fig. 3b). At $\tau \sim 0.1$ ms, the magnitude of the external magnetic field has increased from 0 mT to 70 mT and starts reinitialization of $L$ (Supplementary Fig. 6).

We now turn to stronger terahertz current pulses with $j_0 = 0.4 j_0^{\max}$ and $1.0 j_0^{\max}$, which result in signal dynamics with two key features (Fig. 3b). (1) The peak signal amplitude $S_{\text{MLB}}(\tau)$ saturates around 1 in both cases. (2) $S_{\text{MLB}}(\tau)$ reaches an approximately constant value, i.e., a step, within less than 15 ps after excitation. Importantly, the step persists up to 0.1 ms, when reinitialization starts (Supplementary Fig. 6). The step height $S_{\text{MLB}}(+200 \text{ ps}) - S_{\text{MLB}}(-5 \text{ ps})$ depends on the terahertz field strength in a highly nonlinear manner (Supplementary Fig. 7).

Feature (1) is consistent with Eq. (3) and $\sin^2 \varphi_L \leq 1$. The slight signal overshoot is attributed to the uncertainty of our calibration method. Feature (2) suggests that the local $L$ has relaxed to one of the four equilibrium positions of Fig. 1a. In other words, we have $\varphi_L = -90°, 0°, +90°$ or $180°$, which implies signal contributions of $\sin^2 \varphi_L = 1, 0, 1$ or $0$ in Eq. (3), respectively. Therefore, the signal for $\tau > 20$ ps can be written as

$$S_{\text{MLB}}(\tau) = f_{-90°}(\tau) + f_{+90°}(\tau), \qquad (4)$$

where $f_{-90°}$ and $f_{+90°}$ is the volume fraction of domains with $\varphi_L = -90°$ and $+90°$, respectively. Fig. 3b reveals that $S_{\text{MLB}}(\tau = 200 \text{ ps})$ reaches a value up to 0.6. Consequently, in 60% of the probed volume, the Néel vector has rotated by $\pm 90°$ in response to the strongest terahertz current pulse.

Although the MLB signal [Eqs. (3) and (4)] does not distinguish $\pm L$ domains, we still conclude that $f_{+90°}$ prevails over $f_{-90°}$ or vice versa for the following reasons. First, the SFG micrographs of Fig. 2c,d show that $f_{90°}$ and $f_{-90°}$ differ substantially for bare Mn$_2$Au films, and the same behavior is expected in the Mn$_2$Au|Py system. Second, the magnetization $M$ of the Py layer (Fig. 3a) exhibits a pronounced precession (Supplementary Fig. 7). We ascribe it to a reorientation of $M$ toward the new $L$ direction, driven by the interfacial exchange coupling of $L$ and $M$ (Supplementary Fig. 7). As the precession signal $\propto M_z$ is opposite for domains with $\varphi_L = \pm 90°$, its presence indicates the prevalence of one of the orientations, i.e., $f_{90°} \neq f_{-90°}$. Third, we perform atomistic spin-dynamics simulations that agree well with the observed switching behavior.

**Spin-dynamics simulations.** We model a system of $\sim 10^6$ spins in Mn$_2$Au|Py as classical magnetic moments (Methods) [29]. The dynamics of each moment is described by the stochastic Landau-Lifshitz-Gilbert equation with an effective time-dependent field. The latter arises from the exchange and spin-orbit interactions with the surrounding moments, the quasi-static external magnetic field and the driving terahertz charge current, as well as thermal fluctuations at a temperature of 300 K. To assess the impact of thermal fluctuations, each simulation is repeated 50 times.

Fig. 4a displays the experimentally determined terahertz current density $j = ju_x$. Fig. 4b-d show the resulting simulated deflection $\varphi_L(\tau)$ of the Néel vector, averaged over the Mn$_2$Au thickness, for a current peak amplitude of $j_0/j_0^{\max} = 0.2$ (Fig. 4b), 0.8 (Fig. 4c) and 1.0 (Fig. 4d).

In the linear-response regime, at $j_0 = 0.2 j_0^{\max}$, all runs coincide to very good approximation (Fig. 4b). This behavior indicates that stochastic effects due to thermal spin motion are minor. However, for currents around the switching threshold (Fig. 4c-d), we observe that the primary terahertz pulse induces a branching of the dynamics: $L$ either relaxes back to $\varphi_L = 0°$ or settles at $\varphi_L = -90°$. This branching indicates a more prominent role of stochastic torques close to the maximum of the potential-energy barrier at $\varphi_L = -45°$ (Fig. 1a). Even higher current strengths lead to more complex dynamics including $+90°$-switching (Supplementary Fig. 8), which is expected from the bipolar terahertz waveforms.

To compare the simulations to the experimental signals (Fig. 3b), we calculate the expected MLB signal from each simulated $\varphi_L(\tau)$ by means of Eq. (3) and subsequently average over simulation runs. The resulting signal waveforms $S_{\text{MLB}}(\tau)$ (Fig. 4e) inherit the complex dynamics of $\varphi_L(\tau)$. Remarkably, they are in very good agreement with the measured waveforms (Fig. 3b). In particular, our simulated system begins to switch at roughly the same threshold current as in the experiment.

The transition region from $S_{\text{MLB}} \approx 0$ (no switching) toward $S_{\text{MLB}} \approx 1$ (full switching) with increasing current is slightly narrower in the simulations. This apparent discrepancy can be ascribed to the spatial inhomogeneities of the current density in the real sample (Supplementary Fig. 2), whereas our simulations assume a uniform current distribution. The experimental probe, therefore, averages over regions with different current amplitudes and makes the observed signal a superposition of the signals in Fig. 4e.

To summarize, our spin-dynamics simulations reproduce the general features of the experimental signals very well. Importantly, they show that $L$ rotates largely on a deterministic, non-thermal trajectory for an appreciable range of excitation strengths.

**Discussion and conclusion**

Our spatially and temporally resolved probes and our simulations provide evidence of terahertz-current-induced non-thermal switching of the Néel vector $L$ in the antiferromagnetic metal Mn$_2$Au. We find all attractive features of field-like terahertz NSOT: directional switching as given by the polarity of the driving current and without the need of a static magnetic bias field, long-lived and robust current-induced spin ordering in sample regions with initially zero macroscopic antiferromagnetic order, as well as operation at room temperature, with ultrafast speed and high efficiency.

Remarkably, the switching within $15\,\text{ps}$ is more than one order of magnitude faster than switching of ferromagnets based on damping-like spin-orbit torque [9]. We attribute the short settling time to the $\approx 2\,\text{ps}$ short relaxation time of the resonantly driven uniform in-plane magnon of $Mn_2Au$ [46] and to the NSOT that reverses sign once $\varphi_L$ surpasses 90° [Eq. (1)]. The deposited terahertz-pulse energy density corresponds to $\sim 10^{-2}$ eV per pulse for each Mn site (Methods), which is about one order of magnitude lower than for switching by nanosecond pulses in similar structures [8]. It is further comparable to a variety of approaches, such as terahertz switching in the antiferromagnetic insulator $TmFeO_3$ close to the spin-reorientation temperature [21], all-optical magnetization switching in the ferrimagnetic alloy GdFeCo [54] and picosecond current-pulse switching in ferro- and ferrimagnetic structures [9, 40].

Even though the switching time and deposited terahertz pulse energy of the $Mn_2Au$|Py stack are already highly competitive, we anticipate that they can be reduced by more than one order of magnitude. First, while the Py layer is crucial to our pump-probe experiment (Fig. 3a), it is expected to make the response slower and more energy-demanding relative to a bare $Mn_2Au$ film. Indeed, simulations of such a system indicate that the switching time can be reduced down to the magnon oscillation period of $\sim 1\,\text{ps}$ (Supplementary Fig. 9). At the same time, the threshold current is reduced by a factor of 4, implying a factor of 16 smaller dissipated energy. The deposited terahertz pulse energy can be reduced further by shaping of the vectorial terahertz electric-field transients [29, 30].

Interestingly, according to theory work [29, 55], NSOT in $Mn_2Au$ is dominated by current-induced accumulation of electron orbital rather than spin angular momentum. In this sense, our switching results could be considered as an ultrafast implementation of an essential spin-orbitronic functionality.

Finally, we expect that our terahertz methodology can be transferred to all antiferromagnetic systems that show switching by quasi-DC electric-fields, including conventional [13, 14], chiral [10, 11] and synthetic antiferromagnets [16, 56], as well as altermagnets [12, 15, 24] and magnetoelectric antiferromagnetic insulators [17-19]. Our results open a promising pathway toward on-chip applications of terahertz antiferromagnetic spin-orbitronics. It is fueled by a broad range of material candidates and the engineering of spin-orbit, orbital and magneto-electric torques, resonance frequencies and read-out mechanisms.